\documentclass[superscriptaddress, 10pt,twocolumn,a4paper]{article}
\usepackage{authblk}
\usepackage{hyperref}

\usepackage{graphicx}
\usepackage{float} 
\usepackage{amsmath,bm}
\usepackage{color,soul,xcolor} 
\soulregister\cite7
\usepackage[linesnumbered,ruled,vlined]{algorithm2e}
\usepackage{atbegshi}
\usepackage{subfig}

\SetCommentSty{mycommfont}
\SetKwInput{KwInput}{Input} 
\SetKwInput{KwOutput}{Output}  

\newtheorem{theorem}{Theorem}
\newtheorem{lemma}{Lemma}
\newtheorem{definition}{Definition}

\title{An Algorithm for the Decomposition of Complete Graph into Minimum Number of Edge-disjoint Trees}
\author[1]{{Antika Sinha}\thanks{Corresponding Author: \texttt{antikasinha@gmail.com}}}
\author[2]{Sanjoy Kumar Saha}
\author[3]{{Partha Basuchowdhuri}}

\affil[1]{Department of Computer Science, Asutosh College, Kolkata-700026, India}
 \affil[2]{Department of Computer Science \& Engineering, Jadavpur University, Kolkata-700032, India}
 \affil[3]{School of Mathematical \& Computational Sciences, IACS, Kolkata-700032, India}

\pdfoutput=1

\begin{document}
\maketitle

\begin{abstract}

\noindent
In this work, we study methodical decomposition of an undirected, unweighted  complete graph ($K_n$ of order $n$, size $m$) into minimum number of edge-disjoint trees. We find that $x$, a positive integer, is minimum and $x=\lceil\frac{n}{2}\rceil$ 
as the edge set of $K_n$ is decomposed into edge-disjoint trees of size sequence $M = \{m_1,m_2,...,m_x\}$ where $m_i\le(n-1)$ and $\Sigma_{i=1}^{x} m_i$ = $\frac{n(n-1)}{2}$. For decomposing the edge set of $K_n$ into minimum number of edge-disjoint trees, our proposed algorithm takes total $O(m)$ time.  

\end{abstract}

\section{Introduction}
Graphs are ubiquitous in applications that we can see around us. For example, social media websites have underlying follower-followee networks, which can be represented as graphs structures. Such networks arise due to interactions among the users of the social media motivated by a common objective. Such networks are massive in terms of number of vertices. As a result, many graph-based algorithms do not prove to be scalable for these graphs. On the other hand, special graphs, such as trees, even with similar number of vertices, have better scalability. Therefore, it could be interesting to investigate the problem of decomposing a graph into multiple trees to increase the scalability of the algorithm by reducing the running time complexity.

Prior works suggest different ways of organizing subsets of vertices or edges of a graph. Many problems in graph theory can be represented as a decomposition problem e.g., graph coloring, minimum vertex covering number, etc. 
A decomposition is uniform, if all the decomposed subgraphs are of the same size. These are useful for minimizing the imbalance in the sizes of the resultant subgraphs to ensure each of them handles comparable load e.g., distributing computational tasks evenly across multiple processing units in parallel computing. Arboricity~\cite{nash1964decomposition}, treewidth~\cite{robertson1986graph}, tree number~\cite{ringel1997tree}, spanning tree packing number~\cite{palmer2001spanning}, etc. are often considered as important graph decomposition parameters~\cite{ringel1997tree,gao2018arboricity,cheng2023independent}. In this context, the packing and covering of dense graphs~\cite{bondy2008graph,schwartz2022overview} are active research area of graph theory. The general packing or covering problem in graph theory deals with the maximum or minimum number of graphs, whose edge set (pairwise edge-disjoint) union becomes another graph of larger size.  An alternative approach to such problem is graph factorization~\cite{tutte1961problem}. It is a way of representing a graph as the union of disjoint paths or edge-disjoint components (subgraphs) as e.g. stars~\cite{tarsi1981decomposition}, trees, etc. 

Graph decomposition and graph partition problems have been active areas of research since at least 1960. Starting with the seminal work of Tutte~\cite{tutte1961problem}, Nash-Williams~\cite{nash1961edge} and Beineke~\cite{beineke1964decompositions}, we only mention the part of literature work which is the most relevant to our considered problem. 
During 1961, Tutte~\cite{tutte1961problem} \& Nash-Williams~\cite{nash1961edge} individually studied how many edge-disjoint spanning trees a finite graph can have. Nash-Williams~\cite{nash1964decomposition} (1964) generalized the previous results to forests and provided necessary constraints. 
The problem of graph decomposition into minimum number of edge disjoint trees, from a theoretical standpoint, shares a connection with the $k$-tree partition problem~\cite{biedl2007partitions} in which edge set of an input graph is partitioned into $k$ number of edge-disjoint trees. Alternatively, one can represent the edge set of the graph being colored by $k$ colors, where each color represents a tree. Biedl \& Brandenburg established the NP-hardness of this optimization problem for general graphs with $k\ge2$.

 Among different types of graph decomposition methods studied and reviewed in graph theory literature, star decomposition is a well studied graph decomposition problem that describes a graph as union of line/edge disjoint star graphs. In this context, researchers have explored the decomposition of  complete graphs into stars~\cite{cain1974decomposition}, subsequently into paths and stars~\cite{shyu2010decomposition}, etc. 
 These (acyclic decomposition of a graph) are useful in applications like scheduling, load balancing~\cite{gao2014arboricity}, optimal file organization scheme with respect to lesser redundancy,  parallel computing e.g. Master-Slave paradigm in grid or P2P infrastructure where master is responsible to allocate different tasks on the slaves and collect individual results to produce the final result, etc. Besides general graph, considerable attention has been paid to the graph decomposition problem where the input graph is a complete graph. In this regard, from application point of view, Yamamoto et al. in~\cite{yamamoto1975claw} studied the problem of decomposition of a complete graph into a union of line disjoint claws or stars towards designing a balanced file organization scheme. Abueida \& Daven in~\cite{abueida2004multidecompositions} examined variations to the problems of subgraph packing, covering and factorization for  establishing an effective way to decompose large complete graph into   reasonably well-behaved subgraphs. The authors further commented that the general multi-decomposition problem may take an interesting turn if copy of one or both the non-isomorphic subgraphs obtained from the decomposition is a spanning subgraph. Later, Bryant in~\cite{bryant2010packing} established the necessary conditions for packing pairwise edge-disjoint paths of arbitrary specified lengths for complete graph. Recently, Sethuraman \& Murugan in~\cite{sethuraman2021decomposition} have proposed a new conjecture to decompose a complete graph into copies of two arbitrary trees and also discussed related open problems. In the realm of computer networking and distributed systems, problems are frequently represented in terms of graph theory. Graph theoretic methods offer a powerful framework for understanding and addressing challenges inherent in these systems. Thus graph decomposition problem is motivated by the desire to partition networks into subsets with specific properties, such as being acyclic or containing a defined number of nodes or cycles of certain sizes. This restructuring has the potential to enable system administrators to devise strategies for enhancing performance, improving fault tolerance, etc.

Although various types of graph decomposition exist, we here focus on graph decomposition problem concerning its edge set; particularly, in the context of decomposing edge set of an undirected complete graph ($K_n$) into minimum number of edge-disjoint trees, such that union of the decomposed tree-edges constitute the original edge set of $K_n$ with size $\binom n2$, for $n>2$ whereas the cases $n=1$ and $2$ is trivial.

\section{Problem Formulation}
Given an undirected graph $G$ of order $n$ and size $m$, let $V(G)$ and $E(G)$ denote its vertex set and edge set, respectively. 
We start the section by restating two well-known theorems from Tutte and Nash-Williams on decomposition of finite graphs into edge-disjoint spanning trees. Both Tutte and Nash-William  independently investigated Theorem~\ref{theorem:1} which was later generalized for forests by Nash-Williams, as Theorem~\ref{theorem:2}.

\begin{theorem} (Tutte~\cite{tutte1961problem} \& Nash-Williams~\cite{nash1961edge}, 1961):
Let $G$ be a graph and $\kappa$ be a positive integer. Then $G$ contains $\kappa$ edge-disjoint spanning trees if and only if 
\begin{equation}
 |E_P(G)| \ge \kappa(|P|-1)
 \label{eq:tuttenash}
\end{equation}
 \label{theorem:1}
 
  holds for every partition $P$ of $V(G)$ and $E_P(G)$ is the set of the edges of $G$ joining vertices belonging to different members of $P$. Therefore, $G_P$ denotes a graph of vertex set P and edge set $E_P$(G).
\end{theorem}

\begin{theorem} (Nash-Williams~\cite{nash1964decomposition}, 1964): Let $G$ be a graph and $\kappa$ be a positive integer. Then edge set of $G$ is decomposable into $\kappa$ forests if and only if $G$ is sparse, i.e.,
\begin{equation}
 |E(X)| \le \kappa(|X|-1)
 \label{eq:nash}
\end{equation}
where $X \subset V(G), X \neq \emptyset$ and $|E(X)|$ is the number of edges joining elements of $X$. 
  \label{theorem:2}
\end{theorem}

Now we outline three definitions that are relevant for this study.

\begin{definition}
 Spanning Tree Packing number ($\sigma$): Spanning Tree Packing (STP) number, denoted by $\sigma(G)$, is the maximum number of edge-disjoint spanning trees contained in $G$. Palmer~\cite{palmer2001spanning} further investigate STP number for several important families of graphs.
 \label{def:stp}
\end{definition}

\begin{table}[ht]
    \caption{ STP number ($\sigma$) for some useful families of graph from~\cite{palmer2001spanning}}
    \label{tab:stpbounds}
   \begin{center}
    \begin{tabular}{|c|c|} 
    \hline
      \textbf{Graph(G)} & \bm{$\sigma$}\textbf{(G)} \\
      \hline
		$K_n$ $(n\ge1)$ & $\lfloor n/2 \rfloor$   \\
 		\hline	 
		$K_{n_1,n_2}$ $(1\le n_1 \le n_2)$ & $\Bigl\lfloor{ \frac{n_1n_2}{n_1+n_2-1} } \Bigl\rfloor$   \\
 		\hline	 
	   Maximal Planar & $2$   \\
 		\hline	 
    \end{tabular}
  \end{center}
\end{table}

\begin{definition}
 Arboricity ($\alpha$): Arboricity~\cite{nash1964decomposition} of a  graph $G$, denoted as $\alpha(G)$, is defined as the minimum number of edge-disjoint forests that could decompose edge set of $G$. It can be formulated as
 \label{def:arb}
 \begin{equation}
  \alpha(G) = \Bigl\lceil{\frac{|E(G)|}{|V(G)|-1}}\Bigr\rceil = \Bigl\lceil{\frac{m}{n-1}}\Bigr\rceil
   \label{eq:arb}
 \end{equation}
so that $G$ is sparse if decomposable into forests.
 \end{definition}

Arboricity ($\alpha$) and STP number ($\sigma$) are both used to measure the connectivity structure of a graph, but they are conceptually distinct.
The difference between these two graph parameters  lies in the lack of connectivity. Graph arboricity represents how well-connected the graph is, with lower arboricity indicating higher connectivity. On the other hand, STP number ensures extraction of spanning trees (therefore connected), while arboricity does not impose such a restriction on connectivity.  Theorem~\ref{theorem:1} provides a useful lower bound for $\sigma$, which matches well with $\alpha$, if the input graph is densely connected (for example, an undirected complete graph). This motivates us to study graph edge set decomposition into minimum number of edge-disjoint trees, which may not always be spanning trees. Therefore, to reduce the constraints imposed by the earlier definitions, we look into tree covering number~\cite{artes2014tree}.

\begin{definition}
 Tree Covering number ($\tau$): Tree Covering number of a graph $G$ denoted as $\tau(G)$ is the smallest number of edge-disjoint trees to cover the edge set of $G$. $\tau$ can be formulated as    
 \label{def:tcn}
 \begin{equation}
  \tau(G) = min\{|T|\}
 \end{equation}
 \end{definition}
where $T$ = $\{G_1,G_2,G_3,...,G_p\}$ represent a collection of acyclic 
subgraphs of $G$. Then, $T$ is considered a tree cover of $G$ if for every edge $e \in E(G)$, there exists $G_i \in T$ such that $e \in E(G_i)$ for $i=1,2,...,p$. Therefore, in terms of definition, the parameter $\tau$ serves as a bridge linking $\sigma$ and $\alpha$, which is relevant to the problem statement concerning the presently considered problem.

 \subsection{Problem Statement}
Given a finite, undirected, unweighted complete graph ($K_n$) with vertex set $V(K_n)$, edge set $E(K_n)$ of order $n$ and size $m=\binom n2$, find 
a tree set (cover) $T$ containing minimum ($\tau$-many) number of pairwise edge-disjoint trees. 
Mathematically, $E(K_n)$ is to be decomposed into $\tau$ many edge-disjoint trees as $T[1:\tau]$ with size sequence $M=\{m_1,m_2,...,m_{\tau}\}$, such that 
$\Sigma_{i=1}^{\tau} m_i$ = $\frac{n(n-1)}{2}$.

Next we discuss three lemmas that are instrumental in addressing our considered problem of decomposing the complete graph ($K_n$) into the minimum number of edge-disjoint trees.

\begin{lemma}
Let $n,m$ be the order, size of $K_n$ and $x$ be a positive integer such that $M=\{m_1,m_2,...,m_x\}$ be a set of positive integers. There exists $x$ pairwise edge-disjoint trees of sizes $m_1,m_2,...,m_x$ in $K_n$ if and only if, $m_i\le(n-1)$ for all $i=1,2,...,x$ and $\Sigma_{i=1}^{x} m_i$ =$\frac{n(n-1)}{2}$.
\label{lemma_1}
\end{lemma}

\noindent
\textbf{Proof.} We see that the condition  $\Sigma_{i=1}^{x} m_i = \frac{n(n-1)}{2}$ where $m_i\le(n-1)$ for $i=1,2,...,x$ is necessary for complete edge set decomposition of $K_n$ into acyclic subgraphs. However, case $n=1$ and $2$ are trivial. Here, the condition, $m_i$ being less equal to $(n-1)$ is sufficient to verify that each of the obtained subgraph is a tree. 

\begin{lemma}
Let $n,m$ be the order, size of $K_n$ and $x$ be a positive integer such that $M=\{m_1,m_2,...,m_x\}$ be a set of positive integers giving $\Sigma_{i=1}^{x} m_i = \frac{n(n-1)}{2}$ for $i=1,2,...,x$. $x$ is minimum, when $|m_i|$ is either $(n-1)$ or $(n-1)/2$ or $(n-2)$.
\label{lemma_2}
\end{lemma}

\noindent
\textbf{Proof.} From the definition of arboricity, this lemma naturally holds true when $n$ is even. If not, then we have two solutions: a) a trivial lower bound of graph arboricity is obtained by dividing the number of edges by $n-1$, as this is the best one can do for covering all graph edges with a set of edge-disjoint spanning trees. When $n$ is odd, $n/2$ is not an integer but $(n-1)$ is a multiple of 2.  Since, $n/2$ and $(n-1)/2$ are only $1/2$ apart and one of them is an integer. Therefore, $(n-1)/2$ is the largest integer less than $n/2$. b) Moreover, the below recurrence relation, 

\begin{equation}
\begin{split}
\binom nk = \binom {n-1}{k-1} + \binom {n-1}{k}
 \label{eq:binomwk}
 \end{split}
\end{equation}

\noindent
for all integers $n,k$ where $n\ge0$ and $1 \le k<n$, gives $\frac{n(n-1)}{2} \cong (n-1)$mod$(n-2)$ when $k=2$. 
Now consider a scenario where $|m_i|$ is either $(n-1)$ or 
$(n-q)$ where $q>2$ and we still obtain a valid solution. But we claim that this is not possible for $x$ to be minimum when $q>2$ and additionally, find that Eq.~\ref{eq:binomwk} leads to,
\begin{equation}
\begin{split}
\frac{n(n-1)}{2}-(n-1) = \frac{(n-1)(n-2)}{2}
 \label{eq_2lem2wk}
 \end{split}
\end{equation}
when $k$ = 2 and $n$ is odd. Thus the lemma is proved.

\begin{lemma} Edge set of $K_n$ can be decomposed into a set of minimum number of edge-disjoint trees of size sequence $M$ consisting of 
\begin{itemize}
\item $\frac{n}{2}$ occurrences of $n-1$, if $n$ is even.
\item $\frac{n-1}{2}$ occurrences of $n-2$ and $1$ occurrence of  $n-1$, if $n$ is odd.
\end{itemize}
\label{lemma_3}
\end{lemma} 

\noindent
 \textbf{Proof.} It can be trivially shown for a complete graph of even order. On the other hand, when $n$ is odd, we can prove this Lemma following Lemma~\ref{lemma_2} and Eq.~\ref{eq_2lem2wk}. Thus we obtain $\tau(K_n)=\lceil n/2 \rceil$.

\section{Algorithm}
In this section, we present our proposed algorithms namely Complete Graph Decomposition: a) DECK-E in Algorithm~\ref{algo:decke} for even order complete graph and b) DECK-O in Algorithm~\ref{algo:decko} for odd order complete graph. Their objective is to decompose the edge set of an undirected, unweighted  complete graph ($K_n$) of order $n$ into minimum number ($\tau$) of edge-disjoint trees as tree set $T[1:\tau]$. 

\begin{figure}[H]
\centering
\framebox{
\includegraphics[scale=1] {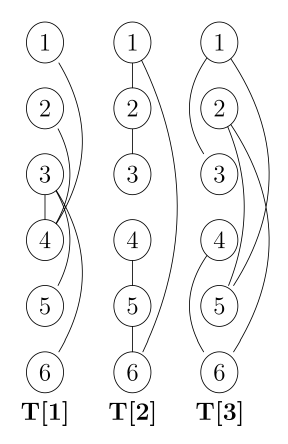}}
\caption{The resulting tree set $T[1:3]$ obtained using algorithm~\ref{algo:decke}: DECK-E where  the input graph is $K_6$.
}
\label{fig:n6}
\end{figure}

Initially, we emphasize the importance of Lemma~\ref{lemma_1}~-~\ref{lemma_3} in the proposed algorithms. Lemma~\ref{lemma_1} structures the considered problem of decomposing a complete graph into the minimum number of edge-disjoint trees (with no edge repetition) in accordance with the definition of arboricity. Lemma~\ref{lemma_2} addresses the potential size sequences of any resulting tree in $T$, derived from arranging the edge set of the input graph $K_n$ into minimum number of edge-disjoint trees. Continuing from this point, Lemma~\ref{lemma_3} gives us the size  sequence to follow for the edge set decomposition of  $K_n$ into the minimum number of edge-disjoint trees, focusing on the what rather than the how. Following that, we illustrate the algorithm steps as a result.

\subsection{Results}
\begin{figure}[H]
\centering
\framebox{
\includegraphics[scale=1.0] {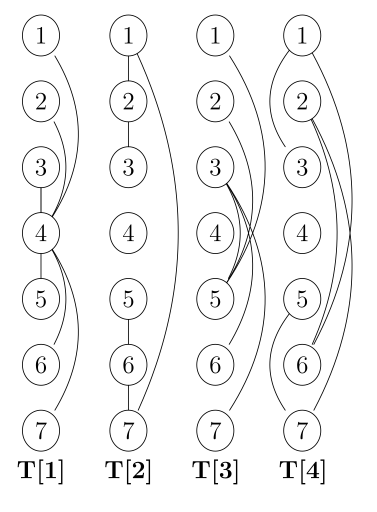}}
\caption{The resulting tree set $T[1:4]$ obtained using algorithm~\ref{algo:decko}: DECK-O where  the input graph is $K_7$. 
}
\label{fig:n7}
\end{figure}

Here, we illustrate the step-by-step construction of the edge set for the resulting edge-disjoint tree set $T$ when our proposed algorithms are individually applied to any input complete graph $K_n$.
Specifically, we have examined and showcased the decomposed results for input graph $K_6$ in Fig.~\ref{fig:n6} and $K_7$ in Fig.~\ref{fig:n7}.

\begin{algorithm}
\DontPrintSemicolon
  \caption{Decomposition of an even-ordered complete graph (DECK-E($K_n$)) into the minimum number of edge-disjoint trees}
  \label{algo:decke}

  \KwInput{$K_n$ where $n$ is even}

\KwOutput{Tree set $T$}
    \SetKw{Continue}{continue}

$\tau \gets mid \gets \lceil n/2 \rceil$\;
$p, q \gets mid, 1+mid$\;

\tcp{\color{blue}{tree set $T[1:\tau]$ initialization}}
\For{$i \leftarrow 1$ \KwTo $\tau$}{
$T[i] \gets \emptyset$
}

\tcp{\color{blue}{tree set $T[1]$ to build as following}}
$T[1] \gets T[1] \cup e(p,q)$\\
\For{$i \leftarrow 1$ \KwTo $p-1$}{
$T[1] \gets T[1] \cup e(i,q)$\\
$T[1] \gets T[1] \cup e(n-i+1,p)$\
    }

\tcp{\color{blue}{tree set $T[2]$ to build as following}}
\For{$i \leftarrow 1$ \KwTo $n$}{
\If{$i \neq p$}{
$T[2] \gets T[2] \cup e(i,(i+1)\%n)$
}
}
$i \gets 1$ \;

$pt \gets p-i, qt \gets q+i$ \;   

\tcp{\color{blue}{tree set $T[3,...,\tau]$ to build as following}}
\For{$t \leftarrow 3$ \KwTo $\tau$}{
$T[t] \gets T[t] \cup e(pt,qt)$

\For{$j \leftarrow 1$ \KwTo $pt-1$}{
$T[t] \gets T[t] \cup e(j,qt)$\\ 
$T[t] \gets T[t] \cup e(n-j+1,pt)$\;
$j \leftarrow j+1$
    }

\For{$j \leftarrow i$ \KwTo $1$}{
$T[t] \gets T[t] \cup e(j,pt+j)$\\
$T[t] \gets T[t] \cup e(n-j+1,qt-j)$\;
$j \leftarrow j-1$
    }
$i \leftarrow i+1, pt \leftarrow pt-1, qt \leftarrow qt+1$\;    
}
return $T[1:\tau]$
\end{algorithm}

To effectively demonstrate the steps of both algorithms, the vertices are arranged vertically downward. For better clarity, refer to Figs.~\ref{fig:ne} and~\ref{fig:no} in appendix for larger graph sizes.
Subsequently, we provide a brief demonstration of the sequential construction of the edge set for the resulting edge-disjoint tree set $T$, focusing separately for even order and odd order complete graphs as input.

\begin{algorithm}
\DontPrintSemicolon
  \caption{Decomposition of an odd-ordered complete graph (DECK-O($K_n$)) into the minimum number of edge-disjoint trees}
  \label{algo:decko}

  \KwInput{$K_n$ where $n$ is odd}

\KwOutput{tree set $T$}
    \SetKw{Continue}{continue}

$\tau \gets mid \gets \lceil n/2 \rceil$\;
$p, q \gets mid-1,mid+1$\;    
    
\tcp{\color{blue}{tree set $T[1:\tau]$ initialization}}
\For{$i \leftarrow 1$ \KwTo $\tau$}{
$T[i] \gets \emptyset$
}

\tcp{\color{blue}{tree set $T[1]$ to build as following}}
\For{$i \leftarrow 1$ \KwTo $n$}{
\If{$i \neq mid$}{
$T[1] \gets T[1] \cup e(i,mid)$
    }
}
\tcp{\color{blue}{tree set $T[2]$ to build as following}}
\For{$i \leftarrow 1$ \KwTo $n$}{
\If{$i \neq p$ \textbf{or} $i \neq mid$}{
$T[2] \gets T[2] \cup e(i,(i+1)\%n)$
}
}
$i \gets 0$ \;

$pt \gets p, qt \gets q$ \;   

\tcp{\color{blue}{tree set $T[3,...,\tau]$ to build as following}}
\For{$t \leftarrow 3$ \KwTo $\tau$}{
$T[t] \gets T[t] \cup e(pt,qt)$

\For{$j \leftarrow 1$ \KwTo $pt-1$}{
$T[t] \gets T[t] \cup e(j,qt)$\\ 
$T[t] \gets T[t] \cup e(n-j+1,pt)$\;
$j \leftarrow j+1$
    }

\For{$j \leftarrow i$ \KwTo $1$}{
$T[t] \gets T[t] \cup e(j,pt+j)$\\
$T[t] \gets T[t] \cup e(n-j+1,qt-j)$\;
$j \leftarrow j-1$
    }
$i \leftarrow i+1, pt \leftarrow pt-1, qt \leftarrow qt+1$\;    
}
return $T[1:\tau]$
\end{algorithm}

\noindent
\textbf{$\textbf{n}$ even case:} Here, we elaborate on the steps of our proposed algorithm, denoted as Algorithm~\ref{algo:decke} DECK-E. As depicted in Fig.~\ref{fig:n6}, the application of the algorithm DECK-E to $K_6$ results in the  tree set $T[1:3]$. By leveraging Lemma~\ref{lemma_3} and Table~\ref{tab:stpbounds},  one can readily observe that the edge set of $K_6$  can be decomposed into three non-repeating copies of edge-disjoint spanning trees. Following the initialization of an empty tree set $T$, steps 5 to 8 constructs the first tree $T[1]$, steps 9 to 11 produce the second tree $T[2]$ and finally,  steps 14 to 24 provide the edges of the last tree (or set of trees for complete graphs of even order $K_{n>6}$) $T[3]$  as edge set of $K_6$ is decomposed using algorithm DECK-E. It is worth noting that $K_6$ is the smallest even order complete graph for which  every steps of Algorithm~\ref{algo:decke} is executed at least once.

\noindent
\textbf{$\textbf{n}$ odd case:} Here, we delve into the steps of our proposed algorithm, denoted as Algorithm~\ref{algo:decko} namely DECK-O.  Referring to Lemma~\ref{lemma_3}, we observe that the application of algorithm DECK-O to $K_7$ yields the tree set $T[1:4]$ (with no edge repetition), as depicted in Fig.~\ref{fig:n7}.  Following the initialization of an empty tree set $T$, steps 5 to 7 constructs the first tree $T[1]$, steps 8 to 10 produce the second tree $T[2]$, and finally, steps 13 to 23 iteratively arrange the edges for the remaining trees (or next set of trees for complete graphs of odd order $K_{n>7}$) $T[3]$ and $T[4]$ as the edge set of $K_7$ is decomposed using DECK-O. Note that $K_7$ is the smallest odd order complete graph for which  every steps of Algorithm~\ref{algo:decko} is  executed at least once.

To accomplish the goal, our algorithm takes total $O(m)$ time to process every edge of $K_n$.

\section{Discussion}

Graph decomposition refers to the process of breaking down a graph into smaller, more manageable components. Different graph decomposition methods include tree decomposition, clique decomposition, edge decomposition, and more. Each type of decomposition serves distinct purposes and proves valuable in analyzing and resolving problems within graph theory.
To address diverse applications, the study of graph decomposition, partitioning, packing, and covering problems is a widely explored area in graph theory research~\cite{gao2018arboricity,cheng2023independent,bondy2008graph,schwartz2022overview}. However, the graph decomposition problem shares similarity to the $k$-tree partition problem, which has been shown to be hard for general graphs when $k\ge2$ in~\cite{biedl2007partitions}.

In this study, we have considered the problem of decomposing the edge set of a finite, undirected and unweighted complete graph ($K_n$) of order $n$ into minimum number of edge disjoint trees.  To accomplish the task,  we have studied two well known theorems by Tutte and Nash-William as Theorem~\ref{theorem:1}~-~\ref{theorem:2}, definitions of useful graph decomposition parameters namely STP number ($\sigma$), arboricity ($\alpha$), tree covering number ($\tau$) in  Definition~\ref{def:stp}~-~\ref{def:tcn} and thereafter   formulated Lemma~\ref{lemma_1}~-~\ref{lemma_3} which further lead to our proposed linear-time algorithms as a) Algorithm~\ref{algo:decke}: DECK-E and b) Algorithm~\ref{algo:decko}: DECK-O  for decomposing the edge set of even and odd order $K_n$, respectively. The algorithm steps are then illustrated using examples for both even and odd order inputs.  We have summarized the findings regarding spanning tree packing number ($\sigma$) using Eq.~\ref{eq:tuttenash} and Table~\ref{tab:stpbounds},  arboricity ($\alpha$) using Eqs.~\ref{eq:nash} and~\ref{eq:arb}, tree covering number ($\tau$) using Lemma~\ref{lemma_3} for complete graphs (given upto order $20$) in Table~\ref{tab:etable} and~\ref{tab:otable}. We noticed that our considered problem can be directly mapped to the spanning tree packing problem~\cite{itai1988multi} when the order of the complete graph is even, see Tables~\ref{tab:stpbounds} and~\ref{tab:etable}. However, when the graph order is odd, the result obtained from our approach (using  Lemma~\ref{lemma_3}) that  initially extracts a spanning tree and then distributes the remaining graph edges into $(\tau-1)$ number of edge-disjoint trees, does not surpass the bound of graph arboricity, see Table~\ref{tab:otable}.  We observe that in cases where the graph's density is high, it might be feasible to cover the edge set using the minimum number of trees~\cite{nash1961edge}, or alternatively, forests. However, this does not align with the characteristics of real-world graphs, which tend to be generally sparse~\cite{nash1964decomposition}. \\


\begin{table}
    \caption{ A comparative analysis of important graph decomposition parameters such as the STP number ($\sigma$), arboricity ($\alpha$) and tree covering number ($\tau$) for $K_n$ of even order (upto $n=20$, case $n=2$ is trivial), using Table~\ref{tab:stpbounds}, Eqs.~\ref{def:stp}~-~\ref{def:arb} and Lemma~\ref{lemma_3}}
    \label{tab:etable}
   \begin{center}
    \begin{tabular}{|l|c|c|} 
    \hline
      \textbf{$n$} & \textbf{$m$} &
      \textbf{$\sigma(K_n)=\alpha(K_n)=\tau(K_n)$} \\
      \hline
		4 & 6 & 2   \\
		\hline     
		6 & 15 & 3  \\
		\hline
		8 & 28 & 4   \\
      \hline
		10 & 45 & 5  \\
		\hline     
		12 & 66 & 6  \\
		\hline
		14 & 91 & 7  \\
      \hline
		16 & 120 & 8  \\
		\hline     
		18 & 153 & 9  \\
		\hline
		20 & 190 & 10  \\
 		\hline	   
    \end{tabular}
  \end{center}
\end{table}

\begin{table}
    \caption{ A comparative analysis of important graph decomposition parameters such as the STP number ($\sigma$), arboricity ($\alpha$) and tree covering number ($\tau$) for $K_n$ of odd order (upto $n=20$, case $n=1$ is trivial), using Table~\ref{tab:stpbounds}, Eqs.~\ref{def:stp}~-~\ref{def:arb} and Lemma~\ref{lemma_3} }
    \label{tab:otable}
   \begin{center}
    \begin{tabular}{|l|c|c|c|} 
    \hline
      \textbf{$n$} & \textbf{$m$} &
      \textbf{$\sigma(K_n)$} & \textbf{$\alpha(K_n)=\tau(K_n)$} \\
      \hline
		3 & 3 & 1 & 2  \\
		\hline     
		5 & 10 & 2 & 3  \\
		\hline
		7 & 21 & 3 & 4  \\
      \hline
		9 & 36 & 4 & 5  \\
		\hline     
		11 & 55 & 5 & 6 \\
		\hline
		13 & 78 & 6 & 7 \\
      \hline
		15 & 105 & 7 & 8  \\
		\hline     
		17 & 136 & 8 & 9  \\
		\hline
		19 & 171 & 9 & 10  \\
 		\hline	   
    \end{tabular}
  \end{center}
\end{table}

Besides theoretical significance, our study holds practical value across various applications, including graph drawing, parallel computing, distributed systems, and reliability, among others.
In network theory, the concept of multiple disjoint paths refers to the existence of several routes between a pair of nodes that do not share common edges. 
The availability of multiple disjoint paths enhances fault tolerance, load balancing, and overall network resilience, making them essential in various applications such as routing protocols~\cite{itai1988multi}, network design, fault-tolerant systems, fault tolerance of transportation networks~\cite{lin2020constructing}, etc. 
Although these paths increase the effective bandwidth between pairs of vertices, they also contribute redundancy and resilience to the network, ensuring reliable communication even in the presence of failures or congestion~\cite{sidhu1991finding,medard1999redundant}. 
Our proposed technique offers an efficient method for decomposing the edge set of any undirected and unweighted complete graph into the minimum number of edge-disjoint trees. We believe that our work adds to the existing body of literature on decomposing graph edge sets into the minimum number of trees by presenting a linear-time algorithm for complete graphs. Next it would be interesting to explore the same problem for general graphs as input, which poses both theoretical challenges, given its complexity~\cite{biedl2007partitions} in graph theory, and practical applications in network design, optimization, etc.



\begin{figure*}
\centering
\framebox{
\includegraphics[scale=0.9] {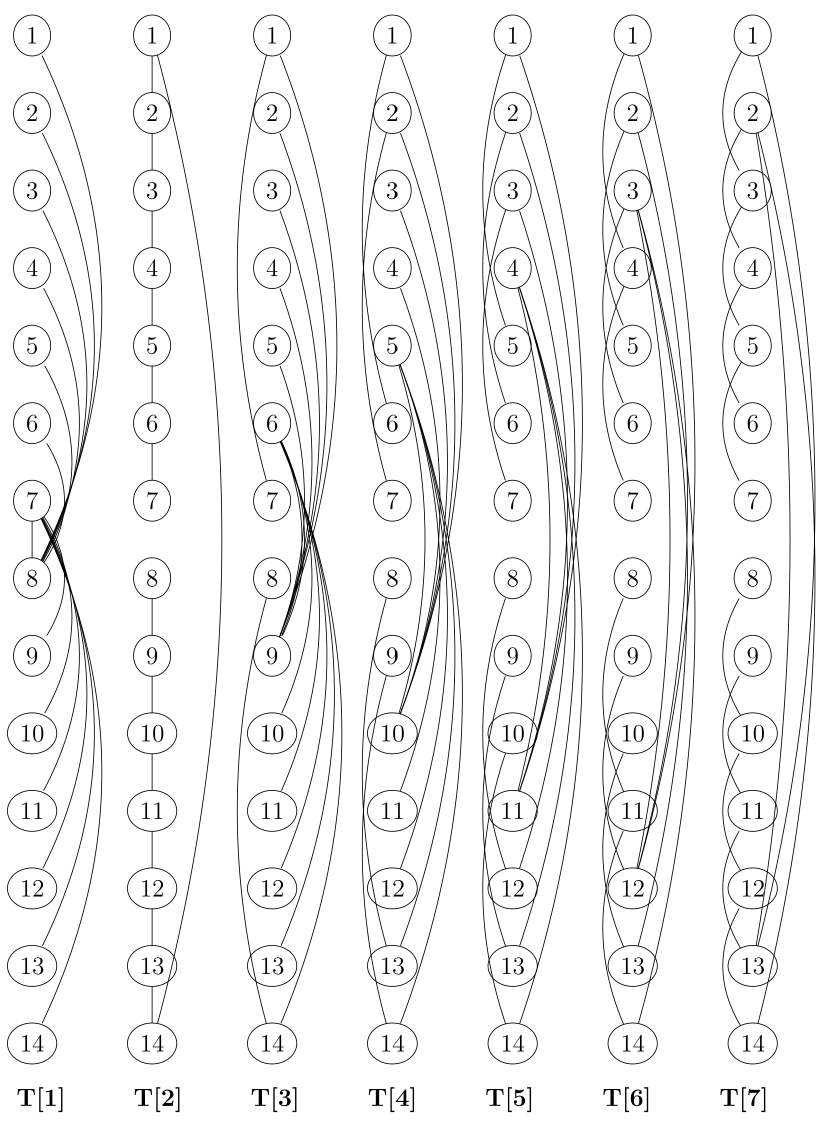}}
\caption{The tree set $T[1:7]$ obtained by decomposing edge set of $K_{n=14}$ into the minimum number ($\tau=7$) of edge-disjoint trees, following our proposed Algorithm~\ref{algo:decke} DECK-E, designed specifically for even values of $n$.
}
\label{fig:ne}
\end{figure*}

\begin{figure*}
\centering
\framebox{
\includegraphics[scale=0.9] {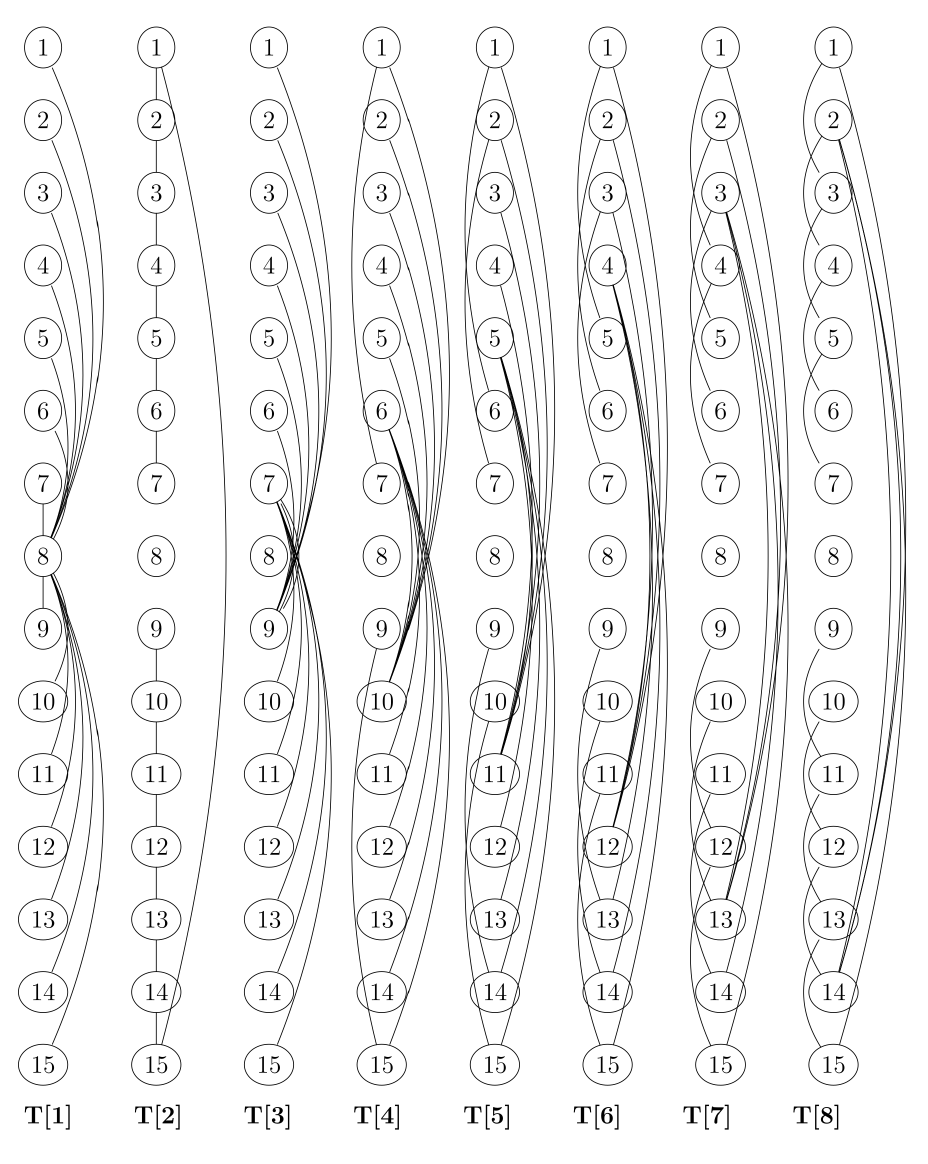}}
\caption{The tree set $T[1:8]$ obtained by decomposing edge set of $K_{n=15}$ into the minimum number ($\tau=8$) of edge-disjoint trees, following our proposed Algorithm~\ref{algo:decko} DECK-O, designed specifically for odd values of $n$. 
}
\label{fig:no}
\end{figure*}

\end{document}